\documentclass[11pt]{article}
\usepackage{amsmath,amssymb,amsfonts}
\usepackage[margin=1in]{geometry}

\title{Geometric approach to quantum theory. L-functionals}
\author{A. Schwarz}
\date{}

\begin{document}
\maketitle
\abstract{This publication consists of slides from my talk at the Simons Center for Geometry and Physics in 2024. It contains a brief review of the L-functional formalism, along with a discussion of its possible applications to QED, linearized gravity, and quenched disorder. The most interesting part is the discussion of the infrared problem in QED and a conjecture on how to construct an infrared-finite perturbation theory for QED.}

\bigskip
 { \bf Geometric approach to quantum theory. L-functionals.}

A. Schwarz, ``Quantum mechanics and quantum field theory from algebraic and geometric viewpoints'', 2024, Springer.

Algebraic approach: starting point is a $*$-algebra $\mathcal{A}$ (unital associative algebra with involution $*$).

Cone of states $\mathcal{C}$ -- linear functionals $\omega \in L$ obeying $\omega(A^*A) \geq 0$ for all $A \in \mathcal{A}$. (Here $L = \mathcal{A}^{\vee}$ -- dual space.)

Set of normalized states $N$: normalization condition $\omega(1) = 1$.

Geometric approach: starting point is a convex set of states $N$ or cone of states $\mathcal{C}$ that are subsets of Banach space $L$. L-functionals.

Evolution operators $T_\tau$ -- automorphisms of $N$.

Decoherence from interactions with adiabatic random perturbations. Derivation of probabilities from decoherence.

Classical theories with a restricted set of observables (our devices allow us to measure only a part of the observables).

Quantum mechanics and its generalizations from such theories.

Geometric theories with a commutative group of time translations and spatial translations $\to$ QFT.

Particles as elementary excitations of the ground state.

Quasiparticles as elementary excitations of translation-invariant stationary state.

Inclusive scattering matrix.

Inclusive cross section = probability density of the process
\[
(M,N) \to (P, Q, \ldots, R) + \text{something}
\]

Can be obtained from matrix elements of inclusive scattering matrix.

Inclusive scattering matrix can be expressed in terms of generalized Green functions on shell (an analog of LSZ formula).

Rediscovered in:

``What can be measured asymptotically?''
S.~Caron-Huot, M.~Giroux, H.S.~Hannesdottir, S.~Mizera,
Journal of High Energy Physics 2024 (1), 1--63.

Asymptotic observables $\approx$ inclusive scattering matrix.

Green function = expectation value of chronological product (times descending),
\[
G_n(x_1, \ldots, x_n) = \omega\big(T(A_1(x_1,t_1)\ldots A_n(x_n,t_n))\big)
\]
$A_i \in \mathcal{A}$, $\omega$ -- translation-invariant stationary state.

Generalized Green functions $\omega(MN)$: $M$ -- chronological product, $N$ -- antichronological product (times ascending).

Appear in the Keldysh formalism of non-equilibrium statistical physics and in the formalism of L-functionals.

If the theory has a particle interpretation, the inclusive scattering matrix carries the same information as the conventional scattering matrix.

However, an inclusive scattering matrix can exist even if the conventional scattering matrix does not exist (for example, for quasiparticles).

Existence theorem for theories with strong cluster property (with a gap).

In QED conventional scattering matrix does not exist (every process involving fixed number of particles has zero probability).

Inclusive scattering matrix in QED can be defined as a limit of inclusive scattering matrices of theories with a gap.

Inclusive cross-sections and inclusive scattering matrix are well defined. Cancellation of infrared divergences.

\bigskip

\textbf{L-functionals} Sch(1967)

Let us quantize a classical theory with finite or infinite number of degrees of freedom.

If $p_k, q^k$ have standard Poisson brackets after quantization, we obtain operators $\hat p_k, \hat q^k$ obeying canonical commutation relations (CCR).

We are working with operators
\[
\hat a(f) = \int dk\, f(k)\hat a(k), \qquad \hat a^+(f) = \int dk\, f(k)\hat a^+(k)
\]
where $f$ runs over the space of test functions $E$ considered as pre-Hilbert space. The integral over $k$ is considered as an integral over continuous parameters and a sum over discrete parameters.

The CCR can be written in the form
\[
[\hat a(f), \hat a(g)] = [\hat a^+(f), \hat a^+(g)] = 0, \qquad [\hat a(f), \hat a^+(\bar g)] = \hbar \langle f, g \rangle
\]
where $f, g \in E$. Assume that $\hbar = 1$.

In the case of an infinite number of degrees of freedom, there exist representations of CCR that are not equivalent to the standard Fock representation where $\hat a(f), \hat a^+(f)$ can be interpreted as annihilation and creation operators (i.e.\ there exists a cyclic vector $\theta$ obeying $\hat a(f)\theta = 0$). Vectors and density matrices in all representation spaces can be regarded as states of the theory at hand.

We can represent the states as functionals
\[
L_K(f) = \mathrm{Tr}\, \hat W_f K,
\]
where $\hat W_f = e^{-\hat a^+(f)} e^{\hat a(\bar f)}$. It is easy to verify that this functional is well-defined for a density matrix $K$ in any representation of CCR.

One can say that when working with functionals $L$ we consider all representations of CCR simultaneously.

Defining a correlation function in the state $K$ as
\[
\mathrm{Tr}\, \hat a^+(k_1)\ldots \hat a^+(k_m) \hat a(l_1) \ldots \hat a(l_n) K
\]
we can say that L-functional is a generating functional of correlation functions.

To emphasize that $L_K$ does not depend analytically on $f$ we use the notation $L_K(\bar f, f)$ or $L_K(f^*, f)$.

Weyl algebra = algebra generated by $a(f), a^+(f)$ obeying CCR.

Exponential form of Weyl algebra = algebra $W$ of operators in Fock space containing all operators of the form $W_f$ and closed in norm topology.

$N$ is the set of normalized positive linear functionals $\sigma$ on $W$ represented by non-linear functionals $\sigma(W_f)$ on $E$.

The space $L$ should be identified with the space of linear functionals on $W$ (dual space) or with the space of non-linear functionals on $E$.

Every element $B$ of algebra $\mathcal{A}$ specifies two operators in dual space $L$ ($\omega(A) \to \omega(AB)$ and $\omega(A) \to \omega(B^*A)$).

Doubling of fields in the formalism of L-functionals.

Green functions in the formalism of L-functionals = generalized Green functions.

\[
\frac{dL}{dt} = HL
\]
$H$ -- ``Hamiltonian.'' It can be expressed in terms of operators $c_i^+, c_i$ where $c_1^+(\bar f)$ is a multiplication operator by $\bar f$, $c_2^+(f)$ is a multiplication operator by $f$, and $c_1(\bar f), c_2(f)$ are variational derivatives with respect to $\bar f$ and $f$.

\[
\hat H = \sum \Gamma_{m,n}(k_1, \ldots, k_m | l_1, \ldots, l_n) \hat a^+(k_1) \ldots \hat a^+(k_m) \hat a(l_1) \ldots \hat a(l_n)
\]
\[
H = H_{\text{left}} - H_{\text{right}} = H(c_1^+(k), c_2^+(k), c_1(k), c_2(k))
\]

Perturbation theory for evolution operator, correlation functions, and generalized Green functions.

\[
\hat H_0 = \sum_k \epsilon(k) a^+(k) \cdot a(k)
\]
\[
H_0 = -i \sum_k \epsilon(k) \big(c_1^+(k) \cdot c_1(k) - c_2^+(k) \cdot c_2(k)\big),
\]
\[
L_n(f^*, f) = e^{-\sum_k f^*(k) n(k) f(k)}
\]
-- stationary states of $H_0$ (quasi-free states).

L-functional corresponding to equilibrium state
\[
L_n(f^*, f) = e^{-\sum_k f^*(k) n(k) f(k)}
\]
where $n(k) = \dfrac{1}{e^{\beta \epsilon(k)} - 1}$.

$U(t,t_0)$ -- evolution operator for $H_0 + V(t)$.
\[
\frac{dU}{dt} = (H_0 + V(t)) U.
\]
\[
S(t,t_0) = e^{-H_0 t} U(t,t_0) e^{H_0 t_0}
\]
evolution operator in interaction picture
\[
\frac{dS}{dt} = \mathcal{V}(t) S
\]
where $\mathcal{V}(t) = e^{-H_0 t} V(t) e^{H_0 t}$.

Standard perturbation theory for $S(t,t_0)$.

The evolution operator for $H_0 + h(at)V$ where $h(0) = 1$, $h(\pm\infty) = 0$ is denoted by $S_a(t,t_0)$.

\textbf{Adiabatic scattering matrix} for the Hamiltonian $\hat H_0 + \hat V$ = evolution operator in interaction picture for the time-dependent Hamiltonian
\[
\hat H(t) = \hat H_0 + h(at) \hat V
\]
When $a \to 0$ then $h(at)$ changes adiabatically (slowly).

Adiabatic scattering matrix in the formalism of L-functionals = evolution operator in interaction picture for the ``Hamiltonian'' $H_0 + h(at) V$.

Scattering matrix and inclusive scattering matrix are limits of adiabatic scattering matrices multiplied by some simple factors (Likhachev, Tyupkin, Sch).

Scattering matrix $\hat S$ can be expressed in terms of the adiabatic scattering matrix in finite volume $\Omega$ in the following way
\[
\hat S = \lim_{a \to 0} \lim_{\Omega \to \infty} \frac{\hat U_{a,\Omega} \hat S_{a,\Omega} \hat U_{a,\Omega}}{\langle \theta | \hat S_{a,\Omega} | \theta \rangle}
\]
where
\[
\hat U_{a,\Omega} = e^{i \sum_k r_{a,\Omega}(k) a^+(k) a(k)},
\]
the limit is understood as the convergence of matrix elements in the sense of generalized functions.

Inclusive scattering matrix $S$ can be represented in the form
\[
S = \lim U_a S_a U_a
\]
where
\[
U_a = e^{i \int dp\, r_a(p) (c_1^+(p) c_1(p) - c_2^+(p) c_2(p))}
\]
and the function $r_a(p)$ is chosen in such a way that one-particle L-functionals are $S$-invariant. Namely, one can take
\[
r_a(p) = \int_{-\infty}^{0} d\tau\, (\epsilon(p, h(a\tau)) - \epsilon(p))
\]
where $\epsilon(p,g)$ are one-particle energies of the Hamiltonian $\hat H(g) = \hat H(0) + g \hat V$.

\bigskip

{\it Relation between inclusive scattering matrix $S$ and conventional scattering matrix $\hat S$}
\[
S L_K = L_{\hat S K \hat S^*}
\]

The limit
\[
\lim_{a \to 0} S_a(0, -\infty) L_n(f^*, f) = \lim_{a \to 0} S_a(0, -\infty) e^{-\sum_k f^*(k) n(k) f(k)}
\]
is a stationary state of $H_0 + V$ (an equilibrium state if we started with equilibrium state of $H_0$). Diagram technique for correlation functions for this state in the formalism of L-functionals coincides with diagram techniques in Keldysh formalism and TFD.

\textbf{ Quenched disorder.} Some coefficients in $\hat H$ are random. We should make calculations for any choice of coefficients and take an average. In the non-stationary case we can calculate the evolution of state in perturbation theory and take an average with respect to random coefficients; this can be done in the language of L-functionals or in Keldysh formalism.

In the stationary case one takes an average of correlation functions with respect to random parameters for fixed temperature. If $T = 0$ we can use L-functionals to calculate the average. In general, L-functionals allow us to calculate the average for fixed entropy of equilibrium state.

\bigskip

\textbf{Example}

\[
\hat V = \sum_k \big(J(k) \hat a^+(k) + J^*(k) \hat a(k)\big)
\]
\[
V = -i \sum_k \big(J^*(k) c_1^+(k) + J(k) c_2^+(k)\big)
\]

In the interaction picture
\[
\frac{dL}{dt} = -i \sum_k \big(J(k) e^{i\epsilon(k) t} c_1^+(k) + J^*(k) e^{-i\epsilon(k) t} c_2^+(k)\big) L
\]
\[
S(t,t_0) = e^{-i \sum_k \big(M(k,t) c_1^+(k) + M^*(k,t) c_2^+(k)\big)}
\]
where $\dot M(k,t) = J(k) e^{i\epsilon(k)t}$.

Phonons interacting with impurities.

\bigskip

\textbf{Example}: QED when we neglect the action of photons on electrons (joint work with I.\ Frolov).

Time-dependent Hamiltonian
\[
\hat H = \hat H_0 + \hat V = \int dk\, \epsilon(k) a^+(k) \cdot a(k) + \int \frac{dk}{\sqrt{2\epsilon(k)}} \big(j(k,t) \cdot a^+(k) + j^*(k,t) \cdot a(k)\big)
\]
where $a(k)$ is a vector potential of electromagnetic field with components $a^\mu(k)$, $\mu = 0, \ldots, 3$ satisfying the Lorenz gauge condition $k^\mu a_\mu(k) = 0$. The scalar product of two 4-vectors has the form $p \cdot k = pk - p^0 k^0$. We use the notation $\epsilon(k) = |k|$.

We suppose that $j^\mu(k,t)$ is a numerical function (Fourier transform of divergence-free current).

The definition of L-functional should be modified. We define
\[
L_K(\alpha^*, \alpha) = \mathrm{Tr}\, \exp\Big(-\int dk\, \alpha(k) a^+(k)\Big) \exp\Big(\int dk\, \alpha^*(k) a(k)\Big) K
\]
where $\alpha(k) = (\alpha^\mu(k))$ obey $\alpha^\mu(k) k_\mu = 0$.

To have explicit Lorentz-invariance one should further modify the definition taking
\[
L_K(\alpha^*, \alpha) = \mathrm{Tr}\, \exp\Big(-\int \frac{dk}{\sqrt{2\epsilon(k)}} \alpha(k) a^+(k)\Big) \exp\Big(\int \frac{dk}{\sqrt{2\epsilon(k)}} \alpha^*(k) a(k)\Big) K
\]

There is no necessity to modify the definition of L-functional in Coulomb gauge (radiation gauge).

The equation of motion for L-functional corresponding to this Hamiltonian has the form $dL/dt = HL$ where $H = H_0 + V$ and
\[
H_0 = -i \int dk\, \epsilon(k) \big(c_1^+(k) \cdot c_1(k) - c_2^+(k) \cdot c_2(k)\big),
\]
\[
V = -i \int \frac{dk}{\sqrt{2\epsilon(k)}} \big(j(k,t) \cdot c_1^+(k) + j^*(k,t) \cdot c_2^+(k)\big)
\]

In the interaction picture we obtain the following equation for the evolution operator $S(t)$
\[
\frac{dS}{dt} = \mathcal{V} S(t) \quad \text{where}
\]
\[
\mathcal{V} = -i \int \frac{dk}{\sqrt{2\epsilon(k)}} \big(e^{i\epsilon(k)t} j(k,t) \cdot c_1^+(k) + e^{-i\epsilon(k)t} j^*(k,t) \cdot c_2^+(k)\big)
\]

A solution of this equation can be found in the form
\[
e^{\sum_{i=1}^{2} M_i(t) \cdot c_i^+}
\]

We obtain
\[
L(\alpha^*, \alpha, t) = \exp\bigg(\int_{t_0}^{t} d\tau \int \frac{dk}{\sqrt{2\epsilon(k)}} \big(e^{i\epsilon(k)\tau} j(k,\tau) \cdot \alpha^*(k) + e^{-i\epsilon(k)\tau} j^*(k,\tau) \cdot \alpha(k)\big)\bigg) L(\alpha^*, \alpha, t_0)
\]

The formula for the solution can be rewritten in the form
\[
L(\alpha^*, \alpha, t) = \exp\bigg(\int dk \sqrt{2\epsilon(k)} \big(e^{-i\epsilon(k)t} A(k,t) \cdot \alpha^*(k) + e^{i\epsilon(k)t} A^*(k,t) \cdot \alpha(k)\big)\bigg) L(\alpha^*, \alpha, t_0)
\]

We use the notation
\[
A^\mu(k,t) = \frac{1}{2\epsilon(k)(2\pi)^{3/2}} \int_{t_0}^{t} d\tau\, e^{i\epsilon(k)(\tau - t)} j^\mu(k,\tau).
\]
where $A^\mu(k,t)$ is the expectation value of electromagnetic potential.

We obtain the inclusive cross-section
\[
dN(k) = A(k,t) \cdot A^*(k,t) \, 2\epsilon(k)\, dk
\]

We calculated the expectation value of the operator
\[
\rho(k) = \sum_{i = \pm} (\varepsilon_i^* \cdot a^+(k))(\varepsilon_i \cdot a(k)),
\]
where $\varepsilon_i$ are polarizations of outgoing photons.

If we are interested in the inclusive cross-section of emission of $n$ photons with momenta $k_1, \ldots, k_n$ then similar calculations lead to the following formula:
\[
dN(k_1, \ldots, k_n) = \prod_{i=1}^{n} A(k_i, t) \cdot A^*(k_i, t)\, 2\epsilon(k_i)\, dk.
\]

\bigskip

\textbf{QED}

\[
S = S_{\text{mat}} + S_{\text{ph}} + \int dx\, j^\mu(x) A_\mu(x)
\]
where $j^\mu = \bar\psi \gamma^\mu \psi$.

In Coulomb gauge $(\mathrm{div}\, A = 0)$ we have
\[
\hat H = \hat H_{\text{mat}} + \hat H_{\text{ph}} - jA + \hat V_{\text{nl}}
\]
where $\hat V_{\text{nl}} = \displaystyle\int dx\, dx' \frac{:\rho(x)\rho(x'):}{8\pi |x - x'|}$ and $\rho(x) = j^0(x)$ stands for charge density operator, or
\[
\hat H = (\hat H_{\text{mat}} + \hat H_{\text{ph}} - j_{\text{num}} A) + (-j + j_{\text{num}}) A + \hat V_{\text{nl}}
\]

In formalism of L-functionals we have doubling of fields.

\textbf{Conjecture.} With the right choice of the numerical part of current there are no infrared divergences in the formalism of L-functionals if in perturbation theory the first line is considered as free Hamiltonian.

\textbf{More precise conjecture.} To get rid of infrared divergences in inclusive cross section of some process we should choose the numerical part of current coinciding with the current of incoming particles at $t \to -\infty$ and with current of outgoing particles at $t \to +\infty$.

\bigskip

\textbf{Sketch of proof}

Represent the Hamiltonian of QED in the form
\[
\hat H_{\text{mat}} + \hat H_{\text{ph}} + \hat U + (-jA - \hat U) + \hat V_{\text{nl}}
\]
where $\hat U = -j_U A$,
\[
j_U(k,t) = \int dp\, \frac{p}{p^0} e^{i \omega_k(p,k) t} \rho(p)
\]
$\rho(p) = \sum_i (b^{i+}(p) b^i(p) - d^{i+}(p) d^i(p))$ is charge density operator.

Then the second line does not contribute to infrared divergences. Applying the formalism of L-functionals and taking the ``Hamiltonian'' corresponding to the first line as free Hamiltonian we get rid of infrared divergences.

\bigskip

\textbf{Infrared divergences from $-jA$}

In the expression $-jA$ represent $j$ as a sum of two terms: first term $j_T$ where every summand contains one creation and one annihilation operator and the second term $j_{T'}$ where every summand contains either two creation or two annihilation operators. The time dependence of the second term is governed by exponent $\exp(\pm i \omega_m(p,k) t)$ where
\[
\omega_m(p,k) = \sqrt{(p+k)^2 + m^2} + \sqrt{p^2 + m^2} \pm k
\]
is bounded below by positive number $2m$. It follows that the corresponding term of the Hamiltonian $T' = -j_{T'} A$ does not contribute to infrared divergences.

The time dependence of the first term $j_T$ is governed by exponent $\exp(i \omega_k(p,k) t)$ where
\[
\omega_k(p,k) = \sqrt{(p+k)^2 + m^2} - \sqrt{p^2 + m^2} \pm k = \frac{pk}{p^0} + O(k^2).
\]
The corresponding term of the Hamiltonian has the form $T = -j_T A = U + U'$ where
\[
j_T^\mu(k) = j_U^\mu(k) + j_{U'}^\mu(k) = \int dp \sum_{i,j} e^{i\omega_k(p,k) t} u^j(p+k) b^{j+}(p+k) \gamma^\mu u^i(p) b^i(p) - \{b, u \to d, v\}
\]
where $b^i(p), d^i(p)$ are annihilation operators of electron and positron and $u^i(p), v^i(p)$ are corresponding wave functions.

$j_{U'}$ can be written in the form
\[
j_{U'}^\mu(k) = \int \frac{dp}{2m} \sum_{i,j=\pm} e^{i\omega_k(p,k)t} \Big( \big(u^j(p+k) b^{j+}(p+k)(2p^\mu + k^\mu) - u^j(p) b^{j+}(p) 2p^\mu\big)
\]
\[
+ i\, u^j(p+k) b^{j+}(p+k) \sigma^{\mu\nu} \frac{k_\nu}{2m} \Big) u^i(p) b^i(p) - \{b, u \to d, v\}
\]

We used the Gordon relation
\[
\bar u(p+k) \gamma^\mu u(p) = \bar u(p+k) \left( \frac{(2p+k)^\mu}{2m} + i \sigma^{\mu\nu} \frac{k_\mu}{2m} \right) u(p)
\]

The operator $U' = -j_{U'} A$ does not contribute to infrared divergences.

\bigskip

\textbf{Example. Linearized gravity in Minkowski space}

\[
g_{\mu\nu} = \eta_{\mu\nu} + h_{\mu\nu}, \qquad |h_{\mu,\nu}| \ll 1
\]
\[
\bar h_{\mu\nu} = h_{\mu\nu} - \frac{1}{2} h \eta_{\mu\nu}, \qquad h = \eta^{\mu\nu} h_{\mu\nu}
\]

Gauge transformations $h_{\mu\nu} \to h_{\mu\nu} + \partial_\mu \xi_\nu - \partial_\nu \xi_\mu$.

Lorenz (harmonic) gauge $\partial^\mu \bar h_{\mu\nu} = 0$.

\[
\Box \bar h_{\mu\nu} = -16\pi T_{\mu\nu}
\]

\[
L_K = \mathrm{Tr}\, e^{i \int dx\, \alpha^{\mu\nu} h_{\mu\nu}} K
\]

True radiation gauge (Chen--Zhu)
\[
\partial^i h^\rho_i - \frac{1}{2} \partial^\rho h^i_i = 0
\]
\[
\Box h_{ij} = \text{``transverse part'' of energy-momentum tensor}
\]

Wave form.

Inclusive cross-section of gravitons.

\bigskip

\end{document}